\documentclass[fleqn,usenatbib,useAMS]{mnras}

\usepackage{graphicx}	
\usepackage{amsmath}	
\usepackage{amssymb}	
\usepackage{multicol}        
\usepackage{bm}		
\usepackage{pdflscape}	

\usepackage[T1]{fontenc}
\usepackage{ae,aecompl}

\usepackage{newtxtext,newtxmath}

\title[Density jump of collisionless shocks with anisotropic upstream]{Density jump as a function of magnetic field strength for parallel collisionless shocks with anisotropic upstream pressure}

\author[A. Bret et al.]{
Antoine Bret$^{1,2}$\thanks{E-mail: antoineclaude.bret@uclm.es}
\\
$^{1}$ETSI Industriales, Universidad de Castilla-La Mancha, 13071 Ciudad Real, Spain\\
$^{2}$Instituto de Investigaciones Energ\'{e}ticas y Aplicaciones Industriales, Campus Universitario de Ciudad Real, 13071 Ciudad Real, Spain
}

\date{Last updated -; in original form -}

\pubyear{2023}

\begin{document}
\label{firstpage}
\pagerange{\pageref{firstpage}--\pageref{lastpage}}
\maketitle

\begin{abstract}
The properties of collisionless shocks are frequently assessed in the magnetohydrodynamics (MHD) model. Yet, in a collisionless plasma, an ambient magnetic field can sustain a stable anisotropy in the upstream or the downstream, resulting in a departure from the MHD predicted behavior. We present a model allowing to derive the downstream anisotropy, hence the shock density jump, in terms of the upstream quantities. For simplicity, the case of a parallel shock in pair plasma is considered. Contrary to previous works where the upstream was assumed isotropic, here the upstream anisotropy $A=T_\perp/T_\parallel$ is a free parameter. The strong sonic shock regime is formally identical to the isotropic upstream case. Yet, for intermediate sonic Mach numbers, a variety of behaviors appear as a result of the anisotropy of the upstream.
\end{abstract}





\section{Introduction}
Shock waves are fundamental processes which display a sudden jump of density, temperature, etc. in the medium where they propagate. The amplitude of these jumps can be determined assuming that matter, momentum and energy are conserved before and after the shockwave. When binary collisions are frequent, these conservation laws take a simple form as the pressure upstream and downstream can be considered isotropic. In the absence of such collisions, the possible existence of shock waves was debated up to the early 90s \citep{sagdeev1991}.

The reason for the debate was that not only collisions isotropize flow, but they are also the microscopic agent which mediates the very existence of the shockwave. In this respect it was found that shock waves could exist in the absence of collisions only in a plasma, where collective electromagnetic effects can substitute collisions as an agent mediating the shock \citep{Sagdeev66}. Such shock waves have been called ``collisionless shocks''. Their very existence is now firmly established by \emph{in situ} measurements of the earth bow shock \citep{PRLBow1,PRLBow2}.

Collisionless shocks are ubiquitous in astrophysical environments. Their ability to accelerate particles \citep{Blandford78} makes them good candidates to explain the origin of high energy cosmic rays, fast radio bursts \citep{Zhang2020,SironiPRL2021} or gamma gay bursts \citep{Piran2004}. Because the spectral index of shock accelerated particles depends on the shock density ratio \citep{Blandford78}, this quantity is of prime relevance for the physics of the aforementioned phenomena. As mentioned previously, the earth bow shock is another example of collisionless shock. More generally, shock detections in the solar system are routine \citep{David2022}, and some instances of anisotropic upstream have been reported \citep{Fraschetti2020}.

From the theoretical point of view, the absence of collisions requires dealing with collisionless shocks at the kinetic level, making use of the Vlasov equation. Since this equation is notoriously difficult to address, the properties of collisionless shocks are frequently deduced from magnetohydrodynamics (MHD), even though MHD, as the plasma counterpart of fluid dynamics, assumes collisions, hence, isotropy of pressures (\cite{Goedbloed2010} ch. 2 and 3, \cite{TB2017} \S 13.2).

Therefore, one needs to ask the question: to which extent can pressure be considered isotropic in a collisionless plasma? In the absence of external magnetic field, the Weibel instability ensures isotropy, at least on time scales longer than the inverse growth rate, since it states that an anisotropic collisionless plasma is unstable \citep{Weibel,SilvaPRE2021}. Yet, an external magnetic field can stabilize an anisotropy in a collisionless plasma. This has been beautifully proved by \emph{in situ} measurements in the solar wind \citep{BalePRL2009,MarucaPRL2011,SchlickeiserPRL2011}. The instabilities limiting the anisotropy are the firehose and the mirror instability \citep{Gary1993}. While in the limit of zero external field they impose isotropy, they allow for a window of stable anisotropies in the presence of a magnetic field.

Note that Fig. 1-top of \cite{BalePRL2009} shows than even though the firehose and mirror thresholds constraint the solar wind measurements, most of them have indeed $T_\perp \sim T_\parallel$.  As stated in the caption of the figure, this results ``largely from Coulomb collisions'', which are eventually not completely absent. In fact, the bottom plot of the same figure shows that the ``collisional age'' of such measurements is higher. Within the framework of the present purely collisionless study, Coulomb collisions are absent so that the two collisionless instability thresholds are the only factors constraining the anisotropy.

As a consequence, MHD may not be well adapted to the study of collisionless shock waves in a magnetized environment. Various authors worked out the MHD conservation equations for anisotropic pressures \citep{Hudson1970,Erkaev2000,Double2004,Gerbig2011}. Yet, the degree of anisotropy in the upstream and the downstream is a free parameter in these equations. Granted, the upstream anisotropy may well be considered an input parameter on which depends the downstream quantities. After all, shock waves analysis usually results in an expression of the downstream quantities in terms of the upstream ones. But a theory leaving the downstream anisotropy as a free parameter is incomplete.

Recently, a model was developed allowing to compute the downstream anisotropy in terms of the (isotropic) upstream parameters \citep{BretJPP2018,BretPoP2019,BretJPP2022}. For simplicity, this model considered pair plasmas, where the parallel and perpendicular temperatures are the same for both species. In contrast, in an electron/ion plasma, species can be heated differently at the front crossing \citep{Guo2017,Guo2018}, which would require dealing with four different temperatures (two for the electrons and two for the ions). Note that preliminary results seem to indicate that the model can be applied to electrons/ions plasma, provided the electron mass is replaced by the ion mass in some key parameters \citep{Shalaby2022}.

Since collisionless shocks in space may well have an anisotropic upstream \citep{Vogl2001,BalePRL2009}, the goal of the present article is to extend the model to the case of an anisotropic upstream.

\begin{figure}
\begin{center}
 \includegraphics[width=\columnwidth]{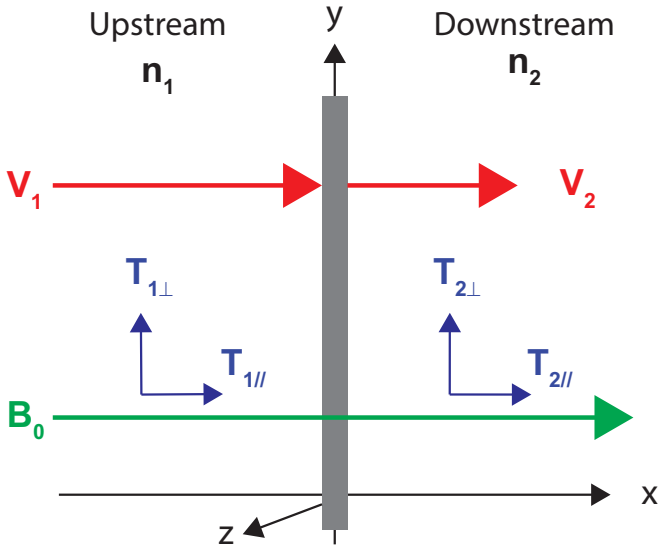}
\end{center}
\caption{System considered. In this parallel configuration, the magnetic field does not change from the upstream to the downstream.}\label{fig:system}
\end{figure}

The system considered is pictured on figure \ref{fig:system}. The upstream and downstream quantities are labelled with subscripts ``1'' and ``2'' respectively. As in \cite{BretJPP2018}, the shock is parallel, with an external field parallel to the direction of motion of the shock. Since in a parallel shock, the upstream and downstream fields are the same \citep{Kulsrud2005}, the magnetic field is labelled $\bmath{B}_0$. There lies the interest of this setup to study collisionless effects effects: because in MHD the fluid and  the field are decoupled for a parallel shock, any change to the shock when the field changes, is a kinetic effect.

For a purely collisionless plasma, the Vlasov equation imposes gyrotropy, namely, equality of the temperatures perpendicular to the field (see \cite{LandauKinetic}, \S 53). This is why we only consider two temperature here instead of three. Admittedly, gyrotropy could be broken, for example, by electric fields developing during the growth of the instabilities involved in this work. Yet, we consider in the sequel the state of the downstream far from the front, and long after the instabilities developed. There and then, the Vlasov equation should enforce gyrotropy.

This article is structured as follows: the method implemented to solve the problem is explained in section \ref{sec:method}. The conservation equations for anisotropic pressures, and the dimensionless variables used here, are presented in section \ref{sec:conser}. In section \ref{sec:upstable} we express the restrictions set on the upstream parameters by the requirement that it be stable. Then from sections \ref{sec:S1} to \ref{sec:together} we solve our model for an anisotropic upstream, before we reach our conclusions in section \ref{sec:conclu}.

\section{Method}\label{sec:method}
The method implemented here is the same than that of \cite{BretJPP2018}. For completeness of the present article, it is now briefly recalled.

If the journey of the plasma through the shock were completely adiabatic, the double adiabatic relations of \cite{CGL1956} would apply,
\begin{eqnarray}\label{eq:CGL}
  T_{2\parallel} &=& T_{1\parallel} \left( \frac{n_2 B_1}{n_1 B_2}  \right)^2, \\
  T_{2\perp} &=& T_{1\perp} \frac{B_1}{B_2}. \nonumber
\end{eqnarray}
For a parallel shock where $B_1=B_2$, they would result in
\begin{eqnarray}\label{eq:CGLpara}
  T_{2\parallel} &=& T_{1\parallel} \left( \frac{n_2}{n_1}  \right)^2, \\
  T_{2\perp} &=& T_{1\perp}. \nonumber
\end{eqnarray}
Yet the shock crossing is not adiabatic since in a shock, there is an entropy increase from the upstream to the downstream. It was then hypothesized in \cite{BretJPP2018} that the relations above hold for the perpendicular temperature, but not for the parallel one. The reason for this is that the plasma compression when crossing the front can be pictured as if it were compressed between 2 converging virtual walls parallel to the shock front, that is, perpendicular to the motion. As a result of such an anisotropic compression, the parallel temperature increases\footnote{By more than the $(n_2/n_1)^2$ factor of Eq. (\ref{eq:CGLpara}).} while the perpendicular one is left unchanged. Note that these heuristic prescriptions have been fully confirmed by PIC simulations in \cite{Haggerty2022}.

This scheme allows to define 2 stages of the kinetic history of the plasma from the upstream to the downstream.

\begin{itemize}
  \item ``Stage 1'' is the state of the plasma right after the front crossing, where its perpendicular temperature has been conserved. As shall be checked in Section \ref{sec:S1}, this prescription, together with the conservation equations, allows for a full characterization of Stage 1.
  \item Depending on its properties, Stage 1 can be stable or unstable. If stable, then it represents the end state of the downstream. If unstable, it migrates to ``Stage 2'', defined as follow.
       \begin{itemize}
         \item If Stage 1 was mirror unstable then Stage 2 lies on the mirror instability threshold.
         \item If Stage 1 was firehose unstable, then Stage 2 lies on the firehose instability threshold.
       \end{itemize}
       Again, in both cases, the conservation equations allow for a full characterization of Stage 2.
\end{itemize}

The shock front is eventually treated as a discontinuity, which means we only consider the state of the plasma far downstream, once it has reached a steady state. This steady state is Stage 1 if it is stable, or Stage 2 if Stage 1 is unstable.

\section{Conservation equations}\label{sec:conser}
The conservation equations for anisotropic pressures were derived in \cite{Hudson1970} and studied in \cite{Erkaev2000,Double2004,Gerbig2011}. For a parallel shock, $m$ being the mass of the particles, they take a simple form,
\begin{eqnarray}
  n_1 V_1                                  &=&  n_2 V_2,         \label{eq:conser1} \\
  n_1 m V_1^2 + P_{\parallel 1}                          &=&  n_2 m V_2^2 + P_{\parallel 2}, \label{eq:conser2} \\
  m\frac{V_1^2}{2} + \frac{P_{\parallel 1}}{n_1} + U_1  &=&  m\frac{V_2^2}{2} + U_2 + \frac{P_{\parallel 2}}{n_2}, \label{eq:conser3}
\end{eqnarray}
where on both sides, the internal energy $U$ is given by
\begin{equation}\label{eq:U}
U=\frac{P_\parallel+2P_\perp}{2n}.
\end{equation}
In Eqs. (\ref{eq:conser2},\ref{eq:conser3}), the pressure term only involves the parallel pressure since it is the pressure component pushing the plasma through the front (see for example the discussion in \cite{FeynmanVol2}, \S 4-3).

The problem is then solved in terms of the dimensionless variables,
\begin{eqnarray}\label{eq:dimless}
  r          &=& \frac{n_2}{n_1}, \nonumber\\
  A_i          &=& \frac{T_{\perp i}}{T_{\parallel i}}, \nonumber\\
  \sigma   &=&  \frac{B_0^2/8\pi}{n_1mV_1^2/2}, \nonumber\\
  \chi_{\parallel 1}^2 &=& \frac{mV_1^2}{P_{\parallel 1}/n_1}.
\end{eqnarray}

The $\chi_{\parallel 1}$ parameter is close to the upstream sonic Mach number. It only misses the adiabatic index $\gamma$. Yet, it is preferable to leave this index out of the definition of any dimensionless number because our model switches continuously from a system with 3 degrees of freedom to a system with 1 degree \citep{BretPoP2021}.

As shall be checked in section \ref{sec:entropyS1}, our model requires $\chi_{\parallel 1} > \sqrt{3}$ for Stage 1 to have a positive entropy jump.

The $\sigma$ parameter is widely used in PIC simulations of collisionless shocks like \cite{Haggerty2022}. It is related to the upstream Alfv\'{e}n Mach number $\mathcal{M}_{A1}$ through
\begin{equation}\label{eq:MA1}
\mathcal{M}_{A1} = \frac{V_1}{c_{A1}} =  \frac{1}{\sqrt{\sigma}},
\end{equation}
where the Alfv\'{e}n speed $c_{A1}$ reads
\begin{equation}
c_{A1} = \sqrt{\frac{B_0^2/4\pi}{n_1m}}.
\end{equation}

\section{Instability restrictions on upstream parameters}\label{sec:upstable}
The range of upstream parameters is restricted by the requirement that the upstream be both firehose and mirror stable. The firehose instability is triggered for an upstream anisotropy $A_1$ fulfilling \citep{Gary1993}\footnote{See \cite{Gary2009,Schlickeiser2010} for the case of pair plasma.}
\begin{equation}\label{eq:firehose1}
A_1 \equiv \frac{T_{\perp 1}}{T_{\parallel 1}} < 1- \frac{1}{\beta_{\parallel 1}},
\end{equation}
with
\begin{equation}\label{eq:beta1}
\beta_{\parallel 1} = \frac{n_1 k_B T_{\parallel 1}}{B_0^2/8\pi},
\end{equation}
where $k_B$ is the Boltzmann constant. The mirror instability is triggered for too strong an upstream anisotropy fulfilling
\begin{equation}\label{eq:mirror1}
A_1   > 1 + \frac{1}{\beta_{\parallel 1}}.
\end{equation}

 Since the upstream has to be stable, Eqs. (\ref{eq:firehose1},\ref{eq:mirror1}) restrict the upstream phase space parameter to consider. In terms of the dimensionless parameters (\ref{eq:dimless}), we find
\begin{equation}
\beta_{\parallel 1} = \frac{2}{\sigma \chi_{\parallel 1}^2},
\end{equation}
so that Eqs. (\ref{eq:firehose1},\ref{eq:mirror1}) read
\begin{equation}\label{eq:upstab}
1 - \frac{\sigma \chi_{\parallel 1}^2}{2} < A_1 < 1 + \frac{\sigma \chi_{\parallel 1}^2}{2}.
\end{equation}

In MHD, the density jump of a parallel shock does not vary with $\sigma$ \cite{Kulsrud2005}. Since the present work eventually aims at studying the deviations from MHD in collisionless shocks, it is advantageous to use $\sigma$ as the main variable and  recast criteria (\ref{eq:upstab}) in terms of it. The upstream is then stable if $\sigma > \sigma_{us}$ (subscript ``us'' for \emph{u}pstream \emph{s}table), with
\begin{equation}\label{eq:sigmaupstab}
  \sigma_{us}  \equiv \left\{
  \begin{array}{l}
   \frac{2}{\chi_{\parallel 1}^2}(1-A_1), ~ \mathrm{when} ~ A_1 < 1 ~(\mathrm{firehose} ~ \mathrm{stability}), \\
  \frac{2}{\chi_{\parallel 1}^2}(A_1-1), ~ \mathrm{when} ~ A_1 > 1 ~(\mathrm{mirror} ~ \mathrm{stability}).
\end{array}
\right.
\end{equation}
For $A_1=1$, any $\sigma$ is allowed since the upstream is always stable.

\section{Analysis of Stage 1, with $T_{\perp 2} = T_{\perp 1}$}\label{sec:S1}
As explained in section \ref{sec:method}, we assume the plasma arrives in the downstream with the same perpendicular temperature as the upstream. This is our Stage 1. It can be characterized by the conservation equations Eqs. (\ref{eq:conser1}-\ref{eq:conser3}). Setting $T_{\perp 2} = T_{\perp 1}$ and  $P=nk_BT$, they read,
\begin{eqnarray}
  n_1 V_1                                  &=&  n_2 V_2,         \label{eq:S1conser1} \\
  n_1 m V_1^2 + n_1k_BT_{\parallel 1}                          &=&  n_2 m V_2^2 + n_2k_BT_{\parallel 2}, \label{eq:S2conser2} \\
  m\frac{V_1^2}{2} + \frac{3k_BT_{\parallel 1}+2k_BT_{\perp 1}}{2}  &=&  m\frac{V_2^2}{2} + \frac{3k_BT_{\parallel 2}+2k_BT_{\perp 1}}{2}. \label{eq:S3conser3}
\end{eqnarray}
Eq. (\ref{eq:S1conser1}) can be used to eliminate $V_2$ everywhere. Then $T_{\parallel 2}$ can be expressed from Eq. (\ref{eq:S2conser2}). Replacing in Eq. (\ref{eq:S3conser3}) gives for $r=n_2/n_1$,
\begin{equation}
(r-1) ( \chi_{\parallel 1}^2(r-2)+3 r)=0.
\end{equation}
The first factor yields the trivial solution $r=1$ corresponding to the absence of any jump. The second factor gives
\begin{equation}\label{eq:rS1}
r = \frac{2 \chi_{\parallel 1}^2}{\chi_{\parallel 1}^2+3},
\end{equation}
formally identical to the isotropic case reported in \cite{BretJPP2018}, replacing $\chi_1$ by $\chi_{\parallel 1}$. Noteworthily, it does not depend on the upstream anisotropy $A_1$.

In the strong sonic shock limit $\chi_{\parallel 1} \rightarrow \infty$, Eq. (\ref{eq:rS1}) gives $r=2$, consistent with a shock in 1 dimension. This is expected since in Stage 1 we freeze the perpendicular degrees of freedom, leaving only 1 free degree, namely, the parallel one.

The downstream anisotropy $A_2$ in Stage 1 is then given by
\begin{equation}\label{eq:A2S1}
A_2 = A_1 \frac{4 \chi_{\parallel 1}^2}{\chi_{\parallel 1}^4+2 \chi_{\parallel 1}^2-3}.
\end{equation}

\begin{figure}
\begin{center}
 \includegraphics[width=\columnwidth]{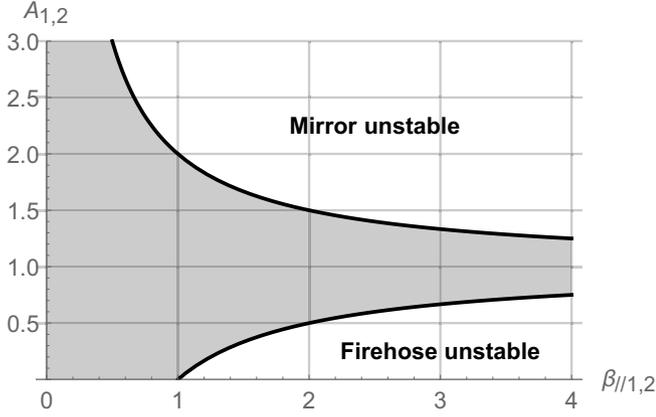}
\end{center}
\caption{Stability diagram for the upstream and the downstream. The plasma is stable within the shaded area.}\label{fig:stab}
\end{figure}

The stability diagram for the firehose and mirror instabilities is pictured in figure \ref{fig:stab}. According to Eq. (\ref{eq:A2S1}), the vertical location of Stage 1 in this diagram only relies on the values of the upstream parameters $A_1$ and $\chi_{\parallel 1}$. Therefore, these are the parameters which determines whether Stage 1 can be mirror or firehose unstable. The ultimate answer to the stability question eventually relies on the field strength $\sigma$, but the choice between these 2 instabilities only relies on $A_1$ and $\chi_{\parallel 1}$. This feature will be key to analyse the end result of our model is section \ref{sec:together}.

\subsection{Entropy of Stage 1 - Restriction on $\chi_{\parallel 1}$}\label{sec:entropyS1}
As is the case in a fluid, the requirement of a positive entropy jump at the shock sets a lower limit on $\chi_{\parallel 1}$. In \cite{BretJPP2018} we found it was $\sqrt{3}$. We shall now see that this lower limit remains the same here.

Expression (\ref{eq:A2S1}) for the anisotropy in Stage 1 allows to compute the entropy jump between the upstream and the downstream in Stage 1.

Following \cite{BretJPP2018}, we consider a bi-Maxwellian of the form
\begin{equation}
f(\bmath{v}) = \frac{n}{\pi^{3/2}\sqrt{a}b}\exp\left( -\frac{v_x^2}{a} \right)\exp\left( -\frac{v_y^2+v_z^2}{b} \right),
\end{equation}
where $a=2k_BT_\parallel/m$ and $b=2k_BT_\perp/m$. The entropy density reads
\begin{equation}
S = -k_B \int f \ln f d^3v = \frac{1}{2}k_B n [3 + \ln (\pi^3ab^2) - 2 \ln n].
\end{equation}
After some straightforward algebra, the difference of entropy per particle $S/n$ between the upstream and the downstream reads
\begin{eqnarray}
\Delta s \equiv \frac{S_2}{n_2}- \frac{S_1}{n_1} &=& \frac{k_B}{2}\ln \left( \frac{A_1}{A_2} \frac{T_{\perp 2}^3}{T_{\perp 1}^3} \right) - 2\ln r \nonumber \\
&=& -\frac{k_B}{2}\ln \left( \frac{A_2}{A_1} r^2 \right),
\end{eqnarray}
since $T_{\perp 2}=T_{\perp 1}$ in Stage 1. From the expression  (\ref{eq:A2S1}) of $A_2$ we find that the entropy jump is the same as for the $A_1=1$ case studied in \cite{BretJPP2018}. Therefore, as in \cite{BretJPP2018}, $\Delta s < 0$ for $\chi_{\parallel 1} < \sqrt{3}$, so that our model is physically meaningful only for $\chi_{\parallel 1} > \sqrt{3}$.

Having characterized Stage 1 we turn to its firehose and mirror stability analysis.

\subsection{Firehose stability of Stage 1}
We now clarify the conditions on $\sigma, A_1, \chi_{\parallel 1}$ for firehose instability of Stage 1. The $\beta_{\parallel 2}$ parameter for the downstream is given by
\begin{equation}\label{eq:beta2S1}
\beta_{\parallel 2} =  A_1 \frac{2r}{\sigma A_2 \chi_{\parallel 1}^2}.
\end{equation}
When substituting $r$ and $A_2$ by their expression (\ref{eq:rS1}) and (\ref{eq:A2S1}), we find
\begin{equation}\label{eq:beta2S1OK}
\beta_{\parallel 2} = \frac{1}{\sigma}\frac{\chi_{\parallel 1}^2 - 1}{ \chi_{\parallel 1}^2},
\end{equation}
identical to the result of \cite{BretJPP2018}.

The criteria for firehose stability reads
\begin{equation}
A_2 > 1 - \frac{1}{\beta_{\parallel 2} },
\end{equation}
that is,
\begin{equation}
\chi_{\parallel 1}^2 \frac{ \sigma (\chi_{\parallel 1}^2+3) + 4 A_1 }{\chi_{\parallel 1}^4+2 \chi_{\parallel 1}^2-3} > 1.
\end{equation}
This equation yields the critical $\sigma\equiv \sigma_{cf}$ below which Stage 1 is firehose unstable,
\begin{equation}\label{eq:sig:cf}
\sigma_{cf} = 1-   \frac{4 A_1}{\chi_{\parallel 1}^2+3}   -   \frac{1}{\chi_{\parallel 1}^2}.
\end{equation}

\subsection{Mirror stability of Stage 1}
The criteria for mirror stability reads
\begin{equation}
A_2 < 1 + \frac{1}{\beta_{\parallel 2} },
\end{equation}
that is,
\begin{equation}
\chi_{\parallel 1}^2 \frac{\sigma (\chi_{\parallel 1}^2+3)-4 A_1}{\chi_{\parallel 1}^4+2 \chi_{\parallel 1}^2-3}  < -1.
\end{equation}
This equation allows to define a critical $\sigma\equiv \sigma_{cm}$ under which Stage 1 is mirror unstable,
\begin{equation}\label{eq:sigcm}
\sigma_{cm} =   \frac{4 A_1}{\chi_{\parallel 1}^2+3}   +   \frac{1}{\chi_{\parallel 1}^2} - 1 = -\sigma_{cf}.
\end{equation}

\begin{figure}
\begin{center}
 \includegraphics[width=\columnwidth]{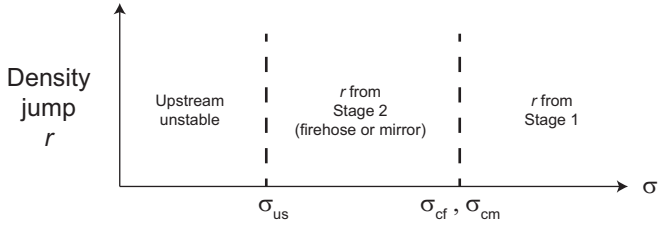}
\end{center}
\caption{Respective position of $\sigma_{us}$, defining the minimum field required to stabilize the upstream, and $\sigma_{cf},\sigma_{cm}$, defining the minimum field required to sustain a stable Stage 1 in the downstream.}\label{fig:sketch_sigma_r}
\end{figure}

\begin{figure}
\begin{center}
 \includegraphics[width=\columnwidth]{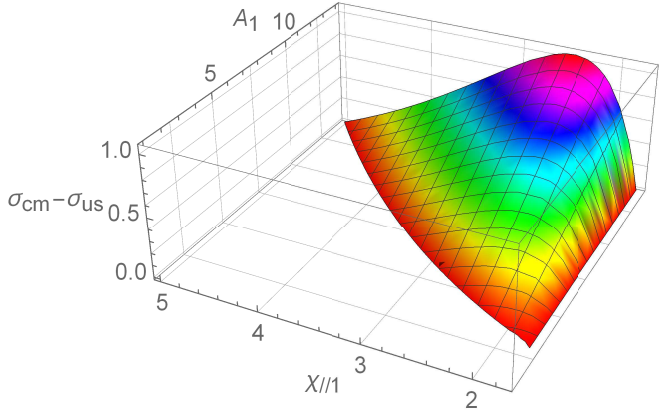}
\end{center}
\caption{Difference $\sigma_{cm}-\sigma_{us}$ in terms of $(\chi_{\parallel 1},A_1)$ where $\sigma_{cm}$ is given by Eq. (\ref{eq:sigcm}) and $\sigma_{us}$ by Eq. (\ref{eq:sigmaupstab}) with $A_1 > 1$.}\label{fig:diffsigma}
\end{figure}

\subsection{Condition for the need of Stage 2}
In the present model,  if Stage 1 is unstable, the downstream switches from Stage 1 to Stage 2.

If Stage 1 is firehose unstable, that is, $A_2 < 1 - \beta_{\parallel 2}^{-1}$, Stage 1 will switch to Stage 2 on the firehose threshold if $\sigma < \sigma_{cf}$.

If Stage 1 is mirror unstable, that is, $A_2 > 1 + \beta_{\parallel 2}^{-1}$, Stage 1 will switch to Stage 2 on the mirror threshold if $\sigma < \sigma_{cm}$.

In addition, the $\sigma$-range is limited by the requirement of a stable upstream, as explained in section \ref{sec:upstable}.

All of these constraints are summarized in figure \ref{fig:sketch_sigma_r}. For Stage 2 to play any role, such a picture obviously requires that the minimum field stabilizing the upstream be smaller than $\sigma_{cf}$ or $\sigma_{cm}$.

For $A_1 < 1$, $\sigma_{cf}$ is relevant and it is easily proved that $\sigma_{us} < \sigma_{cf}$ is always true.

Yet, for $A_1 > 1$, $\sigma_{cm}$ is relevant and $\sigma_{us} < \sigma_{cm}$ is not always fulfilled. Figure \ref{fig:diffsigma} pictures the difference $\sigma_{cm}-\sigma_{us}$ in terms of $(\chi_{\parallel 1},A_1)$. The difference is found positive only for $A_1 > \frac{1}{2}(\chi_{\parallel 1}^2+3)$. As a consequence, for such values of the upstream anisotropy $A_1$, there is no need for Stage 2. An example of such a case will be given in section \ref{sec:together}.

\section{Analysis of Stage 2}\label{sec:S2}
According to the premises of our model, the downstream switches to Stage 2 in case Stage 1 is unstable. From the expression (\ref{eq:A2S1}) of the downstream anisotropy in Stage 1, it is easily proved that $A_1 < 1 \Rightarrow A_2 < 1$. Hence, if $A_1 < 1$, Stage 1 can only be firehose unstable. On the contrary, if $A_1 > 1$, then Stage 1 can only be mirror unstable. We need therefore to study both options, namely, ``Stage 2 firehose'', where the downstream lies on the firehose instability threshold, and ``Stage 2 mirror'', where it lies in the mirror instability threshold.

\subsection{Stage 2 firehose}
Imposing Stage 2 to lie on the firehose instability threshold means imposing $A_2=1-1/\beta_{\parallel 2}$ in the conservation equations (\ref{eq:conser1}-\ref{eq:conser3}). From
\begin{equation}
U_2 = \frac{P_{\parallel 2} + 2 P_{\perp 2}}{2} = \frac{n_2 k_B T_{\parallel 2} + 2 n_2 k_BT_{\perp 2}}{2},
\end{equation}
and
\begin{equation}
A_2 = \frac{T_{\perp 2}}{T_{\parallel 2}} = 1- \frac{B_0^2/8\pi }{n_2 k_B T_{\parallel 2}} ,
\end{equation}
we get
\begin{equation}
U_2 = \frac{1}{2}\left(3k_BT_{\parallel 2} - \frac{B_0^2/4\pi}{n_2}\right).
\end{equation}
Substituting $U_2$ by the expression above in Eq. (\ref{eq:conser3}) and solving the system for the density ratio gives a 2nd order polynomial for the density ratio $r$,
\begin{equation}\label{eq:polyS2fire}
r [(2 A_1+3) r-5]+\chi_{\parallel 1}^2 [r (r+\sigma-5)+4]=0.
\end{equation}
The anisotropy is given by
\begin{equation}
A_2 = \frac{2 r-\chi_{\parallel 1}^2 [r (\sigma-2)+2]}{2 \left[(r-1) \chi_{\parallel 1}^2+r\right]}.
\end{equation}

The polynomial equation (\ref{eq:polyS2fire}) gives 2 solutions. Yet, one of them yields a nonphysical, namely negative, value for the anisotropy $A_2$. The physical density jump that has $A_2>0$ reads
\begin{equation}\label{eq:rS2fire}
rS2_f = \frac{\sqrt{\Delta_f}-(\sigma-5) \chi_{\parallel 1}^2+5}{2 \left(2 A_1+\chi_{\parallel 1}^2+3\right)},
\end{equation}
with
\begin{equation}\label{eq:Deltaf}
\Delta_f = 25-2 \chi_{\parallel 1}^2 (16 A_1+5 \sigma-1)+(\sigma-9) (\sigma-1) \chi_{\parallel 1}^4.
\end{equation}

\subsection{Stage 2 mirror}
Imposing Stage 2 to lie on the mirror instability threshold means imposing $A_2=1+1/\beta_{\parallel 2}$ in the conservation equations (\ref{eq:conser1}-\ref{eq:conser3}). Following the same path than for Stage 2 firehose also gives a 2nd order polynomial for the density ratio $r$,
\begin{equation}\label{eq:polyS2mirror}
\chi_{\parallel 1}^2 [(r (5-r+\sigma)-4)]+r [5-(2 A_1+3) r]=0,
\end{equation}
with the anisotropy
\begin{equation}
A_2 = \frac{\chi_{\parallel 1}^2 (r (\sigma+2)-2)+2 r}{2 \left[(r-1) \chi_{\parallel 1}^2+r\right]}.
\end{equation}

The polynomial equation (\ref{eq:polyS2mirror}) gives 2 solutions. Again, one of them yields a nonphysical, namely negative, value for the anisotropy $A_2$. The physical density jump reads
\begin{equation}\label{eq:rS2mirror}
rS2_m=\frac{\sqrt{\Delta_m}+(\sigma+5)\chi_{\parallel 1}^2+5}{2 \left(2 A_1+\chi_{\parallel 1}^2+3\right)},
\end{equation}
with
\begin{equation}
\Delta_m =  25 - \chi_{\parallel 1}^2 (32 A_1-10 \sigma-2)+(\sigma+9)(\sigma+1) \chi_{\parallel 1}^4.
\end{equation}

\subsection{Some comments on Stage 2}
To which extent can we consider that if Stage 1 is unstable, then Stage 2 remains on the marginal stability threshold, without migrating further into the stability region? As previously stated, \cite{BalePRL2009} concluded that weak Coulomb collisionality is responsible for the concentration of solar wind measurements around $T_\perp = T_\parallel$. Possibly collisionless relaxation \citep{Kadomtsev1970,Lapenta2009,Gedalin2015} could also play a role on longer time scales, which goes beyond the scope of the present work. For the case of an isotropic upstream, the numerical study of \cite{Haggerty2022} found that Stage 2, when triggered by Stage 1 instability, fulfils the marginal stability condition.

While the key assumption for Stage 1 is that the perpendicular temperature is conserved, no such assumption is made on any temperature for Stage 2. There, the only constraint on the downstream is to lie on the firehose or mirror stability threshold. For the case of an isotropic upstream, both temperatures have been explicitly and separately studied  in \cite{BretPoP2021}, showing how in Stage 2, both of them are eventually a function of $\sigma$.

The density jump for Stage 2 is derived from Eqs. (\ref{eq:conser1}-\ref{eq:conser3}). These equations consider the field is homogenous and aligned with the flow, which is why it does not appear in these expressions. Clearly, during the growth of the mirror or the firehose instability, the total field does not have to be homogenous nor aligned with the flow. Yet, as stated in the introduction and at the end of Section \ref{sec:method}, we consider the state of the downstream long after the instabilities developed, once
it has reached a steady state. At this stage, the downstream has reached marginally stability and the macroscopic magnetic field is back in line with the flow. Indeed, the simulations of \cite{Haggerty2022} show that marginal stability is reached within a few $c/\omega_p$'s behind the front\footnote{Where $\omega_p$ is the plasma frequency and $c$ the speed of light.  See for example figure 4(b) of  \cite{Haggerty2022}.}.

\begin{figure}
\begin{center}
 \includegraphics[width=\columnwidth]{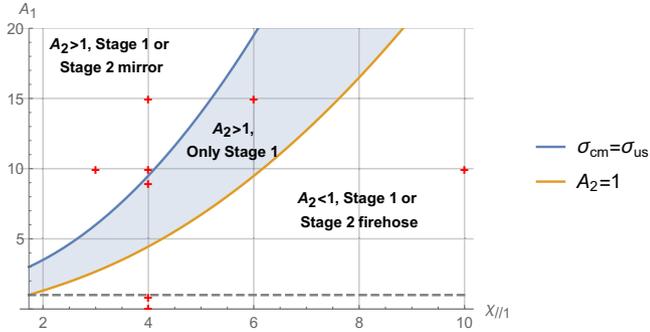}
\end{center}
\caption{Summary of the various cases offered by the model. Upper part: $A_2 > 1$ and the jump is given by Stage 1 or Stage 2 mirror. Intermediate shaded region: the jump can only be given by Stage 1 since $\sigma_{cm} > \sigma_{us}$. Lower part: $A_2 < 1$, the jump is given by Stage 1 or Stage 2 firehose. The red crosses indicate the parameters featuring figure \ref{fig:together}.}\label{fig:summary}
\end{figure}

\section{Putting Stages 1 and 2 together}\label{sec:together}
Having computed all the density jumps for Stage 1 and Stage 2 we can now provided a full picture of the shock density jump in terms of $\sigma,A_1,\chi_{\parallel 1}$.

Figure \ref{fig:summary} pictures the various cases offered by the model in terms of $(A_1,\chi_{\parallel 1})$. In the upper part, $A_2 > 1$ and the jump is given by Stage 1 or Stage 2 mirror. In the intermediate shaded region, the jump can only be given by Stage 1 since $\sigma_{cm} > \sigma_{us}$. In the lower part, $A_2 < 1$, the jump is given by Stage 1 or Stage 2 firehose.

\begin{figure}
\begin{center}
\textbf{(a)}\\
 \includegraphics[width=\columnwidth]{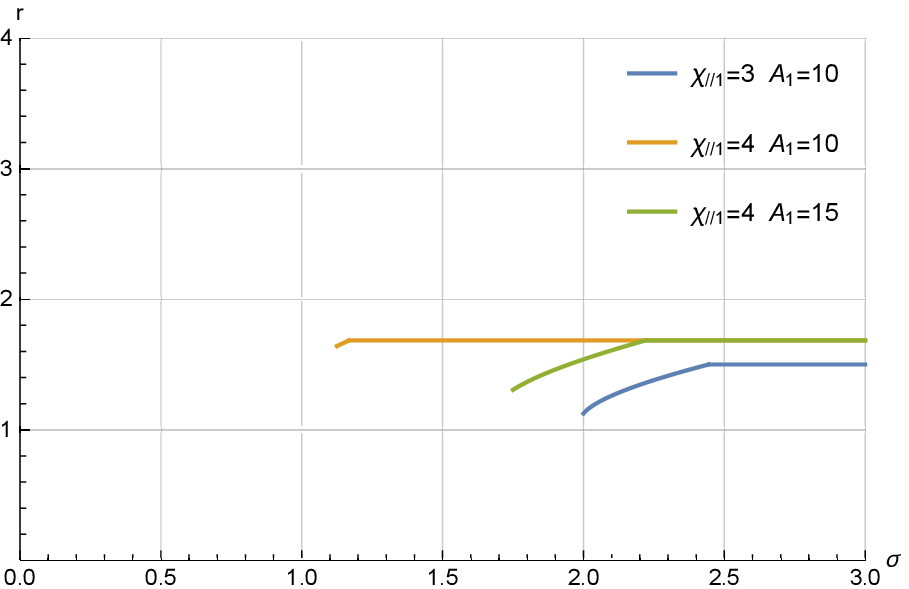}\\
 \textbf{(b)}\\
 \includegraphics[width=\columnwidth]{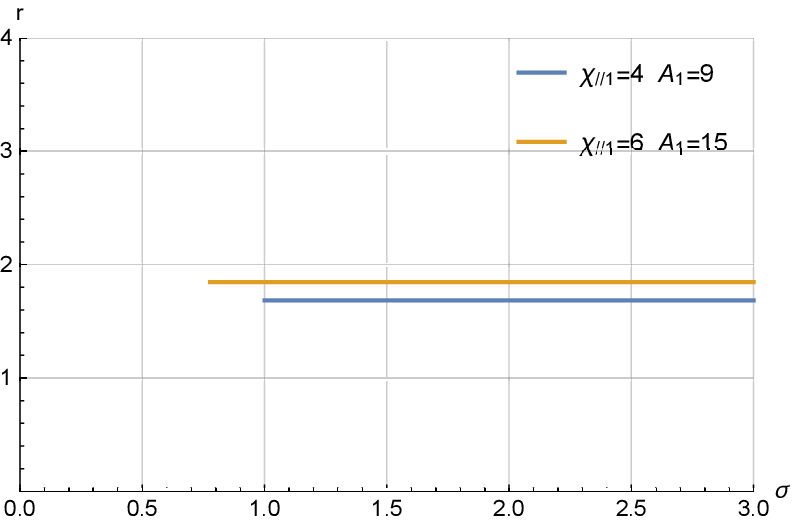}\\
 \textbf{(c)}\\
 \includegraphics[width=\columnwidth]{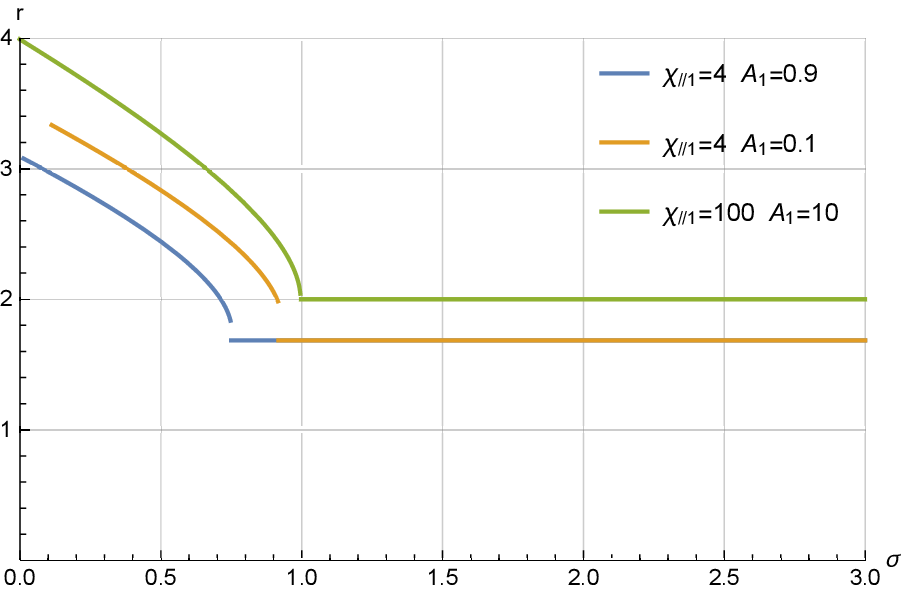}
\end{center}
\caption{Plots $r(\sigma)$ for the values of $\chi_{\parallel 1}$ and $A_1$ marked by the red crosses on figure \ref{fig:summary}.}\label{fig:together}
\end{figure}

Figure \ref{fig:together} presents plots $r(\sigma)$ for the values of $\chi_{\parallel 1}$ and $A_1$ marked by the red crosses on figure \ref{fig:summary}.

Panel (a) shows cases pertaining to the upper part of figure \ref{fig:summary}, with $A_2 > 1$. The curves only start for values of $\sigma$ corresponding to a stable upstream. From Eq. (\ref{eq:sigmaupstab}), this value can be quite high for large anisotropies. The curves present a segment where the jump is the one for Stage 2 mirror [Eq. (\ref{eq:rS2mirror})], since the field is still too weak to stabilize Stage 1. Then the flat part at large $\sigma$ is the jump of Stage 1 [Eq. (\ref{eq:rS1})]. The jump in the Stage 2 mirror phase is an increasing function of $\sigma$. The strong shock case $\chi_{\parallel 1} \rightarrow \infty$ cannot be explored in this regime because, as visible in figure \ref{fig:summary}, this limit always ends up leaving this region of the phase space\footnote{Any point pertaining to the upper left region of figure \ref{fig:summary} (the one labelled ``$A_2>1$, Stage 1 or Stage 2 mirror''), eventually leaves this region if $\chi_{\parallel 1}$ is increased while $A_1$ is kept constant.}.

Panel (b) shows cases pertaining to the shaded part of figure \ref{fig:summary}, where the field required to stabilize the upstream is larger than the minimum field required to mirror stabilize the downstream with $A_2 > 1$. The jump is therefore uniquely given by Stage 1 for all relevant $\sigma$´s. The requirement of a stable upstream still sets a minimum value of $\sigma$ below which a solution would be un-physical.

Finally, panel (c) shows cases pertaining to the lower part of figure \ref{fig:summary}, with $A_2 < 1$. Here, if unstable, Stage 1 is firehose unstable. In this regime, studying the strong shock limit $\chi_{\parallel 1} \rightarrow \infty$ is meaningful since it can be reached without leaving this region. In this respect, $r(\sigma)$ has been plotted for $\chi_{\parallel 1}=100$ (not shown by a red cross on figure \ref{fig:summary}). As evidenced by looking at Eqs. (\ref{eq:rS2fire},\ref{eq:Deltaf}), this limit for Stage 2 firehose cannot depend on $A_1$. In addition, the jump for Stage 1 does not depend on $A_1$ either (see Eq. \ref{eq:rS1}). As a result, the strong shock limit is independent of the upstream anisotropy $A_1$. It is eventually identical to the strong shock result for the isotropic upstream case described in \cite{BretJPP2018}.

\section{Conclusion}\label{sec:conclu}
Since MHD is frequently used to assess the properties of collisionless shocks, it is important to study to which extent this strategy is pertinent \citep{BretApJ2020}. In this respect, parallel shocks are of particular interest since within the MHD framework, the fluid and the field are decoupled \cite{Kulsrud2005}. As a consequence, the density jump of a parallel shock in MHD does  not depend on the field strength.

In a previous work \citep{BretJPP2018}, it was found that due to kinetic effects, the density jump of parallel collisionless shocks can definitely vary with the field strength. For strong enough a field, the density jump can go down to 2, instead of 4, the value predicted by MHD.  The model presented was successfully checked through Particle-In-Cell (PIC) simulations in \cite{Haggerty2022}. Although pair plasmas were considered in these works, preliminary results \citep{Shalaby2022} seem to indicate that the model holds for electron/ion plasmas, provided the ions mass is used to define the dimensionless parameters (\ref{eq:dimless}).

An isotropic upstream has been assumed hitherto. The present work is devoted to extend the model to the case of an anisotropic upstream.

The kinetic history of the plasma as it crosses the shock front is decomposed into 2 stages. First, the perpendicular temperature of the plasma is conserved. This is Stage 1. If it is stable, then this is the end state of the downstream. If Stage 1 is firehose unstable, it migrates to Stage 2 firehose, marginally firehose stable. If  Stage 1 is mirror unstable, it migrates to Stage 2 mirror, marginally mirror stable.

For an isotropic upstream with $A_1=1$, it was found that Stage 1 can only be firehose unstable. Considering $A_1 \neq 1$ brings the possibility that Stage 1 be also mirror unstable. It also restricts the minimum field strength to be considered since the upstream has to be both mirror and firehose stable. The analysis splits the phase space parameters into 3 regions pictured in figure \ref{fig:summary}.

The upper part of figure \ref{fig:summary} has the downstream anisotropy $A_2$ always larger than unity. There, for moderate values of the field, the jump is given by Stage 2 mirror. Then, as soon as the field is strong enough to stabilize Stage 1, the jump is the one of Stage 1. Figure \ref{fig:together}(a) pictures some curves $r(\sigma)$ in this region.

In the middle part of figure \ref{fig:summary}, namely the shaded area, the downstream anisotropy $A_2$ is always larger than unity. Therefore, if unstable, Stage 1 can only be mirror unstable. Yet, in this region, the smallest field required to stabilize the upstream is larger than the smallest field required to mirror-stabilize Stage 1. As a consequence, Stage 2 never occurs in this region. Figure \ref{fig:together}(b) pictures some curves $r(\sigma)$ in this region.

Finally,  in the lower part of figure \ref{fig:summary}, the downstream anisotropy $A_2$ is always lower than unity. If unstable, Stage 1 can only be firehose unstable. In many respects, this case is formally identical to the isotropic case : at large $\sigma$, namely $\sigma > \sigma_{cf}$, the jump is given by Eq. (\ref{eq:rS1}), identical to the isotropic result. For $\sigma < \sigma_{cf}$, the field is too weak to stabilize Stage 1 and the jump is given by Eq. (\ref{eq:rS2fire}, still quite close to the $A_1=1$ case since here, $A_1 \in [0,1]$. In addition, the critical sigma $\sigma_{cf}$ is also formally identical to its $A_1=1$ value\footnote{Compare the present Eq. (\ref{eq:sig:cf}) with Eq. (2.11) of \cite{BretJPP2018}.}. The strong sonic shock limit $\chi_{\parallel 1}\rightarrow\infty$ necessarily pertains to this case, since this region is the only one of figure  \ref{fig:summary} from which one cannot escape when taking this limit. Figure \ref{fig:together}(c) pictures some curves $r(\sigma)$ in this region.

As explained in section \ref{sec:entropyS1}, the present model requires $\chi_{\parallel 1} > \sqrt{3}$ to provide a positive entropy jump. From the definition of $\chi_{\parallel 1}$ given by Eq. (\ref{eq:dimless}), we find $\mathcal{M}_{\parallel 1} = \chi_{\parallel 1} \gamma_\parallel^{-1/2}$, where $\gamma_\parallel$ is the parallel adiabatic index
 and $\mathcal{M}_{\parallel 1}$ the parallel upstream sonic Mach number. Note that in an anisotropic collisionless plasma, the adiabatic index is 3 in the direction parallel to the field, and 2 in the perpendicular directions \citep{Abraham1967}. Hence, the model can describe shocks with  $\mathcal{M}_{\parallel 1} > 1$, for a parallel adiabatic index $\gamma_\parallel=3$.

Future works should contemplate testing these results through PIC simulations, like those of \cite{BretJPP2018} have been tested in \cite{Haggerty2022}.

\section*{Acknowledgements}
A.B. acknowledges support by Grant  PID2021-125550OB-I00 from the Spanish Ministerio de Ciencia e Innovación. Thanks are due to Federico Fraschetti for valuable inputs.

\section*{Data Availability}
The calculations presented in this manuscript were performed using \emph{Mathematica}. The Notebook files will be shared on
reasonable request to the corresponding author.



\bibliographystyle{mnras}
\bibliography{BibBret}






\bsp	
\label{lastpage}
\end{document}